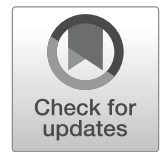

# The use of fast molecular descriptors and artificial neural networks approach in organochlorine compounds electron ionization mass spectra classification

Maciej Przybyłek[1] · Waldemar Studziński[2] · Alicja Gackowska[2] · Jerzy Gaca[2]

**Abstract**
Developing of theoretical tools can be very helpful for supporting new pollutant detection. Nowadays, a combination of mass spectrometry and chromatographic techniques are the most basic environmental monitoring methods. In this paper, two organochlorine compound mass spectra classification systems were proposed. The classification models were developed within the framework of artificial neural networks (ANNs) and fast 1D and 2D molecular descriptor calculations. Based on the intensities of two characteristic MS peaks, namely [M] and [M-35], two classification criterions were proposed. According to criterion I, class 1 comprises [M] signals with the intensity higher than 800 NIST units, while class 2 consists of signals with the intensity lower or equal than 800. According to criterion II, class 1 consists of [M-35] signals with the intensity higher than 100, while signals with the intensity lower or equal than 100 belong to class 2. As a result of ANNs learning stage, five models for both classification criterions were generated. The external model validation showed that all ANNs are characterized by high predicting power; however, criterion I-based ANNs are much more accurate and therefore are more suitable for analytical purposes. In order to obtain another confirmation, selected ANNs were tested against additional dataset comprising popular sunscreen agents disinfection by-products reported in previous works.

## Introduction

Chlorine-containing organic compounds are probably one of the most commonly reported environmental pollutants causing serious problems from decades. There are several major sources of these species including industrial sewage and municipal wastewater (Lee et al. 2006; Antoniou et al. 2006; Sánchez-Avila et al. 2009), pesticides (Karlsson et al. 2000; Carvalho 2017; Harmouche-Karaki et al. 2018; Salvarani et al. 2018; Nambirajan et al. 2018), combustion gases (Morton and Pollak 1987; Hu et al. 2010), or water disinfection by-products (Richardson 2003; Kawaguchi et al. 2005; Moradi et al. 2010). Organochlorine compounds have been frequently detected in surface (Chen et al. 2011; Navarrete et al. 2018; Ali et al. 2018), ground (Shukla et al. 2006; Jayashree and Vasudevan 2007; Chaza et al. 2018) and potable waters (Aydin and Yurdun 1999; Gelover et al. 2000; Palmer et al. 2011), soil (Fang et al. 2017; Thiombane et al. 2018), wastewater, sewage sludge (Bester 2005; Clarke et al. 2010), and marine organisms (Smalling et al. 2010; Gonul et al. 2018; Luellen et al. 2018). Interestingly, there are also natural, non-anthropogenic sources of these compounds such as higher plants, ferns, certain fungi, algae, and phytoplankton

✉ Maciej Przybyłek
  m.przybylek@cm.umk.pl

[1] Chair and Department of Physical Chemistry, Pharmacy Faculty, Collegium Medicum of Bydgoszcz, Nicolaus Copernicus University in Toruń, Kurpińskiego 5, 85-950 Bydgoszcz, Poland

[2] Faculty of Chemical Technology and Engineering, University of Technology and Life Science, Seminaryjna 3, 85-326 Bydgoszcz, Poland

(Gschwend et al. 1985; Engvild 1986; Harper et al. 1988; Wuosmaa and Hager 1990; Gribble 1996).

It has been shown that a number of chloroorganic pollutants exhibit carcinogenic and mutagenic potential causing irreversible damage to living organisms (Lampi et al. 1992; Høyer et al. 1998; Ghosh et al. 2018). These persistent organic pollutants are accumulated in fats and are resistant to biodegradation (Lee et al. 2014). Numerous studies showed that emerging pollutants, such as personal care products or drugs, can enter the environment and undergo conversion under water disinfection conditions to toxic organochlorine compounds (Boorman 1999; Hrudey 2009; Zhao et al. 2010; Hu et al. 2017; Manasfi et al. 2017; Gackowska et al. 2018). In order to evaluate the environmental risk posed by new chlorine-containing pollutants, it is important to use relatively fast and accurate methods of their identification. However, the choice of the method is dependent on the type of the sample. One of the most widely used techniques is GC or HPLC chromatography combined with mass spectroscopy (MS) techniques. Analytical procedures developed for organochlorine pesticide detection deserves special attention. Since pesticides are volatile and thermally stable compounds, gas chromatography and mass spectrometry or tandem mass spectrometry (MS/MS) are commonly used to identify this group of compounds in complex environmental samples. These techniques are particularly useful for the simultaneous detection of compounds with different physicochemical properties (Domínguez et al. 2016). There are many interesting applications of chromatographic methods utilizing mass spectrometry methods. As it was reported in several studies, ultra-high performance liquid chromatography (UHPLC) combined with quadrupole time-of-flight (TOF) mass spectrometer was found to be an efficient and accurate approach for complex wastewater matrices containing pharmaceutics and their metabolites, mycotoxins, and pesticides (Petrovic and Barceló 2006; Martínez Bueno et al. 2007; Ibáñez et al. 2009; Masiá et al. 2014; Jacox et al. 2017). Another interesting examples of advanced methods are techniques combining linear ion trap Orbitrap analyzers with chromatography (Bijlsma et al. 2013; Chen et al. 2017), gas chromatography tandem mass spectrometry (GC-MS/MS) (Raina and Hall 2008; Feo et al. 2011; Barón et al. 2014; Luo et al. 2018; Wang et al. 2018), and liquid chromatography coupled to high resolution mass spectrometry (LC-HR-MS) (Aceña et al. 2015; Kruve 2018). It should be noted, however, that high resolution spectrometers are relatively expensive both to purchase and operate. Besides, these methods require a complex validation processes, and hence are not widely used. Another technique used to determine organochlorine compounds is gas chromatography coupled with selective detectors such as electron capture detector (ECD) (Surma-Zadora and Grochowalski 2008; Dąbrowski 2018), flame photometric detector (FPD), and nitrogen phosphorous detector (NPD). However, they are not appropriate for the simultaneous analysis of a wide range of chloroorganic pollutants. For these reasons, simple mass spectrometry (MS) is still commonly used. As it was reported, the application of efficient isolation methods such as pressurized liquid extraction (PLE) and solid-phase extraction (SPE) along with GC/MS enables for detection of a wide range of chloroorganic pesticides and polychlorinated biphenyls in soil and sediments (Dąbrowski et al. 2002; Dąbrowska et al. 2003). Furthermore, combination of simple liquid-liquid extraction with GC/MS was successfully used for popular sunscreen agents 2-ethylhexyl-4-methoxycinnamate (EHMC) and 2-ethylhexyl 4-(dimethylamino)benzoate (ODPABA) disinfection by-products detection (Nakajima et al. 2009; Santos et al. 2012; Gackowska et al. 2014, 2016; Studziński et al. 2017).

The development of mass spectral interpretation, including spectra prediction, classification, and new fragmentation rules, provides helpful tools for organic compounds identification. This is particularly relevant in case of environmental monitoring comprising detection of analytes in complex matrices. Noteworthy, in many cases, there are no reference standards and no reference spectra available in the literature. There have been several attempts to use theoretical models for EI-MS spectra analysis (Gray et al. 1980; Gasteiger et al. 1992; Copeland et al. 2012; Ásgeirsson et al. 2017; Spackman et al. 2018). According to our best knowledge, 1D and 2D descriptor-based models devoted to the organochlorine compounds have never been reported in the literature. This approach appears to be attractive due to the low computational cost. Recently, many studies have demonstrated that constitutional and topological molecular indices can be successfully applied for predicting different physicochemical properties and biological activities (Duchowicz et al. 2017; Cysewski and Przybyłek 2017; Toropov et al. 2018; Przybyłek and Cysewski 2018). In this paper, a new approach of organochlorine compounds' MS spectra classification was proposed and the aim is to develop computationally efficient and reliable predictive models using fast QSPR/QSAR descriptors and ANNs methodology. Based on this approach, one can confirm the reliability of proposed hypothetical structure by verification of class membership determined using ANNs. Additionally, the analysis of descriptors appearing in the model enables the assessment of the molecular features relevant for the fragmentation behavior of organochlorines.

## Methods

### Mass spectra selection for ANNs' binary classification models generation

The mass spectra data were obtained from NIST database (NIST Chemistry WebBook 2018). The list of compounds along with corresponding [M] and [M-35] peak intensities is provided in online resource S1 (Table S1). The dataset consists of chlorinated hydrocarbons and oxygen-, sulfur-, nitrogen-, and phosphorus-containing organochlorine compounds. Additionally, a different collection comprising disinfection by-products of several sunscreen agents was used as second external test set for models with the highest predicting power.

### Molecular descriptors calculation

Firstly, the IUPAC International Chemical Identifiers (InChIKeys) corresponding to each MS spectra data records were obtained from NIST database. Then, the SMILES codes were generated from InChIKeys with an aid of PubChem Identifier Exchange Service (https://pubchem.ncbi.nlm.nih.gov/idexchange). Finally, these data were used for molecular descriptor calculation taking advantage from PaDEL-Descriptor software (Yap 2011). This was performed using default computation settings.

### Artificial neural network designing and statistical analysis details

All classification models were generated and statistically analyzed using STATISTICA 12 Software (Statsoft, USA). In this study, multilayer perceptron (MLP) algorithm was used and default dataset splitting settings, i.e., 70% for training set, 15% for validation set, and 15% for test set. Training and validation sets are the collections of data used for model generation and its improvement during learning procedure. Test set is the external data collection which was randomly excluded prior to the model generation.

Among 1444 1D and 2D descriptors calculated using PaDEL-Descriptor, only those variables having significant information content, i.e., parameters computable for all molecules and which variance is higher than 0.001, were included. As a result of this analysis, 1056 relevant descriptors were selected. However, this number of variables is still too large to build a reasonable network. In order to avoid overfitting problem, only descriptors with potentially the highest predicting power were used for creating the final models according to preliminary sensitivity analysis approach (Baczek et al. 2004; Mendyk and Jachowicz 2005; Grossi et al. 2007; Cutore et al. 2008; Tirelli and Pessani 2009; Olaya-Marín et al. 2013; Yadav et al. 2014; Song et al. 2015; Rouchier et al. 2016). Therefore, the following procedure was applied. Firstly, five preliminary ANNs involving all 1056 descriptors as input variables were generated automatically. Then, these networks were used for ranking descriptors based on their predicting power. As a result of this step, 100 descriptors with the highest sensitivity were selected, which comprises only 4.5% of the number of considered MS spectra peaks in training set. At the next stage, learning procedure was repeated for selected variables. As a result of this step, for each classification criterion, five ANNs were generated and saved as PMML files (online resource S2).

## Results and discussion

### Characteristics of MS spectra classification models

In case of majority organochlorine compounds, two characteristic MS peaks can be distinguished, namely, molecular ion peak [M] and [M-35] signal which is related to the most abundant chlorine isotope $^{35}Cl$ elimination (Krupčík et al. 1976; Österberg and Lindström 1985; Webster and Birkholz 1985; Nolte et al. 1993; Beil et al. 1997; Pollmann et al. 2001). When [M] is not the base peak, fragmentation proceeds rapidly. On the other hand, high intensity of [M-35] peak denotes relatively high stability of dechlorination products. In this paper, the following two classification criterions were examined and tested against their analytical applicability:

- Criterion I: class 1 ($n = 1588$) comprises [M] signals with the intensity higher than 800 NIST units (according to NIST database the intensity of base peak is 9999), while class 2 ($n = 1599$) contains signals with the intensity lower or equal than 800
- Criterion II: class 1 ($n = 1592$) comprises [M-35] signals with the intensity higher than 100, while [M-35] signals with the intensity lower or equal than 100 belong to the class 2 ($n = 1595$)

By dividing the population in these ways, two large and comparable subsets for each class are obtained. This is important from the statistical viewpoint, since both classes are well represented. The names of the compounds considered in this study along with the classes assigned to them are summarized in online resource S1, Table S1.

The majority of molecular peaks assigned to class 1 can be observed on the MS spectra recorded for aromatic compounds. This seems to be understandable, since π-conjugation enhances the stability of chemical species including ion radicals formed prior to the molecules fragmentation. However, in case of sterically hindered compounds, e.g., 2-chlorotoluene, 3,4-dichlorotoluene, and 2-chloro-1,4-dimethylbenzene (online resource S1, Table S1), the intensity of [M] peak is much lower than [M-35]. This indicates that molecular ion undergoes dechlorination readily. Noteworthy, in case of sterically hindered aliphatic compounds such as 1-hydroxychlordene, 1,4,5,6,7,7-hexachlorobicyclo[2.2.1]hept-5-ene-2,3-dicarboxylic acid, 1,2-dichlorohexane, trichlorfon, 1,1-dichlorocyclohexane, and 1,1,1,5-tetrachloropentane, there are no molecular peaks on the EI-MS spectra. This means that, due to the low stability of molecular ions, the fragmentation proceeds very fast. The influence of steric hindrance on rapid fragmentation has been well documented by many studies (Grützmacher and Tolkien 1977; Shukla et al. 2003; Henderson et al. 2009; Li et al. 2009; Demarque et al. 2016). The absence of molecular peak was observed for 784 compounds of dataset (supplementary Table S1, online resource S1). Some examples are bis(chloromethyl)ether, α,α-Dichloromethyl methyl ether, and carbon tetrachloride.

The brief characteristics of generated networks (ANNs' architecture, learning algorithm and applied error, and activation functions) is summarized in Table 1. In case of all networks, Broyden-Fletcher-Goldfarb-Shanno (BFGS) learning algorithm was applied which is a very popular tool in solving non-linear optimization problems, due to their reliability and good effectiveness (Li et al. 2018). During the learning procedure, the accuracy of the neural network is being gradually improved. Therefore, error function plays an important role. The two types of error functions were applied in the models, sum of squares and entropy. These functions are necessary for modifying neural nets' weights during learning procedure by evaluating the prediction quality of models at particular step (Bishop 1995). Another key features characterizing ANNs are activation functions. The exponential function was found to be the most frequently appearing in case of both hidden and output layers (Table 1).

As we can see from Table 1, in case of all networks representing criterion I and II classification systems, the overall prediction quality which includes both classes is high. However in case of criterion I, exceptionally good accuracy was achieved. Therefore, these models are the most useful from the analytical application perspectives. Testing procedure showed that MLP 100-19-2 ANN is characterized by the highest predicting power. Among 228 mass spectra belonging to class 1, 204 were classified properly (true positives). A slightly better result was achieved for class 2 (237 true positives and 13 false positives).

The relationships between sensitivity (true positive rate) and specificity (true negative rate) can be illustrated by the receiver operating characteristics (ROC) plots. An exemplary ROC charts were summarized on Fig. 1. The ROC plot can be quantitatively characterized using area under the curve (AUC) parameter (Bradley 1997; Mandrekar 2010; Hajian-Tilaki 2013). In case of perfect prediction, the AUC is 1. When AUC is near to 0.5, the quality of the model is poor. In case of criterion I, the AUC values range from 0.9898 to 0.9973 for training set and from 0.9557 to 0.9636 for validation set, which indicates good data fitting achieved during learning procedure. However, the quality of prediction can be evaluated based on the analysis of test set examples, which were excluded prior to the model generation procedure. The AUC values determined for this collection are also very high in case of all ANNs, since they range from 0.9477 to 0.9709. An additional insight into the models' characteristics is provided by the gain plots. On Fig. 2, the cumulative gain plots for the most accurate criterion I-based model (MLP 100-19-2) were presented. As one can see, these plots are typical for good quality binary classification models. Gain charts illustrate the relationship between classified by the model cases and the percentage of true positives. For instance, if we chose half of the compounds assigned by the MLP 100-19-2 model to class 1, more than 90% will be properly classified.

Considering the environmental relevance, several interesting groups of pollutants can be distinguished in the test set. An important class are polychlorinated biphenyls (PCBs). The test set contains 18 PCBs including compounds containing two (PCB 4, PCB 8), three (PCB 33), four (PCB 66, PCB 77, PCB 42, PCB 40, PCB 79), five (PCB 84, PCB 92, PCB

**Table 1** Selected details of created ANN models. In the parentheses the percentages of properly assigned spectra corresponding to class 1 and 2 were presented

| ANN | Learning algorithm | Error function | Activation function | | Model accuracy [%] | | |
|---|---|---|---|---|---|---|---|
| | | | Hidden layer | Output layer | Training | Testing | Validation |
| Criterion I ([M] peak classification models) | | | | | | | |
| MLP 100-19-2 | BFGS 135 | Sum of squares | Exponential | Exponential | 96.68 (96.62; 96.75) | 92.26 (89.47; 94.80) | 92.47 (90.30; 94.61) |
| MLP 100-23-2 | BFGS 129 | Sum of squares | Exponential | Exponential | 96.73 (96.88; 96.57) | 91.84 (89.91; 93.60) | 92.05 (90.72; 93.36) |
| MLP 100-15-2 | BFGS 47 | Entropy | Tanh | Softmax | 98.74 (98.49; 99.01) | 92.05 (92.98; 91.20) | 92.47 (91.56; 93.36) |
| MLP 100-25-2 | BFGS 105 | Sum of squares | Exponential | Linear | 97.09 (96.53; 97.65) | 91.42 (90.35; 92.40) | 92.05 (89.45; 94.61) |
| MLP 100-21-2 | BFGS 43 | Entropy | Tanh | Softmax | 97.62 (97.69; 97.56) | 91.00 (88.16; 93.60) | 92.89 (92.41; 93.36) |
| Criterion II ([M-35] peak classification models) | | | | | | | |
| MLP 100-25-2 | BFGS 36 | Sum of squares | Tanh | Logistic | 91.08 (91.84; 90.30) | 86.19 (86.30; 86.10) | 83.68 (87.40; 79.74) |
| MLP 100-22-2 | BFGS 73 | Sum of squares | Exponential | Linear | 90.86 (91.39; 90.31) | 85.98 (86.30; 85.71) | 86.19 (87.40; 84.91) |
| MLP 100-22-2 | BFGS 72 | Sum of squares | Exponential | Exponential | 88.79 (89.53; 88.04) | 85.98 (85.39; 86.49) | 86.61 (88.62; 84.48) |
| MLP 100-22-2 | BFGS 48 | Sum of squares | Tanh | Linear | 86.96 (86.78; 87.14) | 84.94 (83.56; 86.10) | 86.19 (88.62; 83.62) |
| MLP 100-24-2 | BFGS 65 | Sum of squares | Exponential | Linear | 89.02 (88.00; 89.85) | 85.36 (84.02; 86.49) | 85.98 (86.99; 84.91) |

 

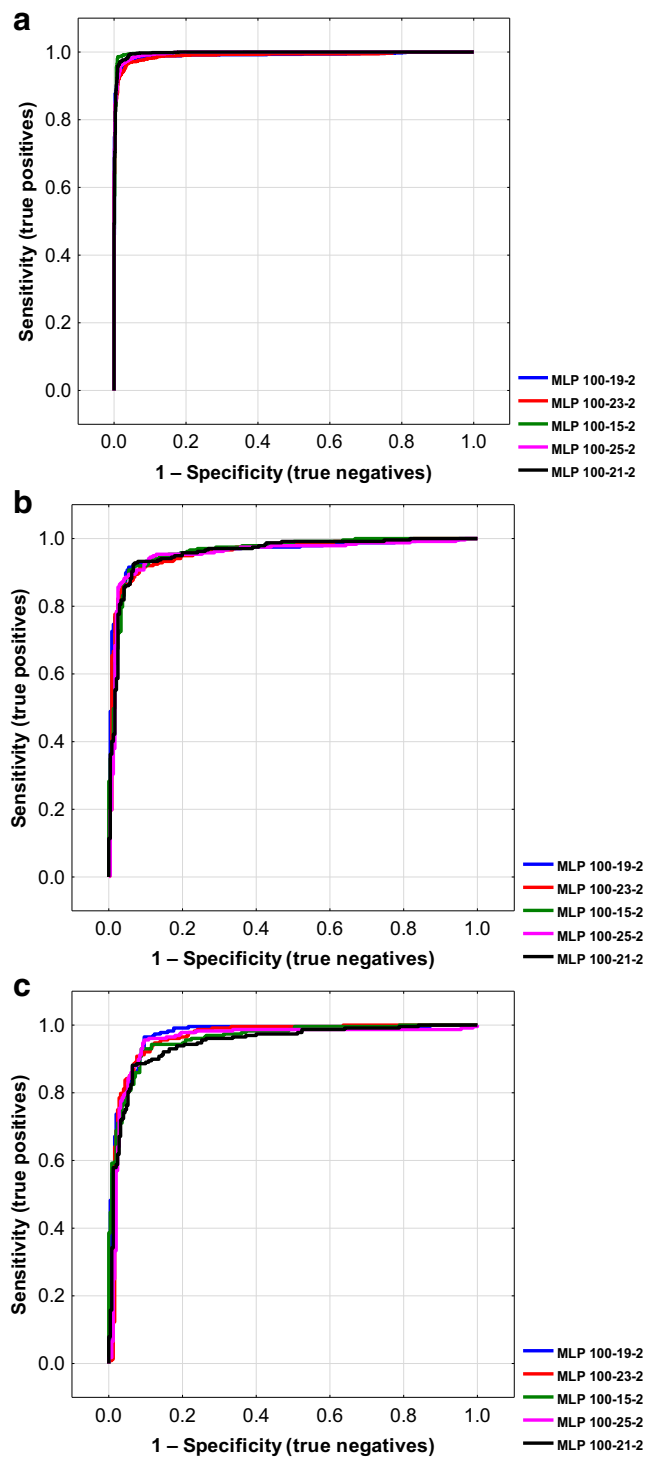

**Fig. 1** Receiver operating characteristic (ROC) plots for training (**a**), validation (**b**), and test sets (**c**) of [M] peak classification models (criterion I)

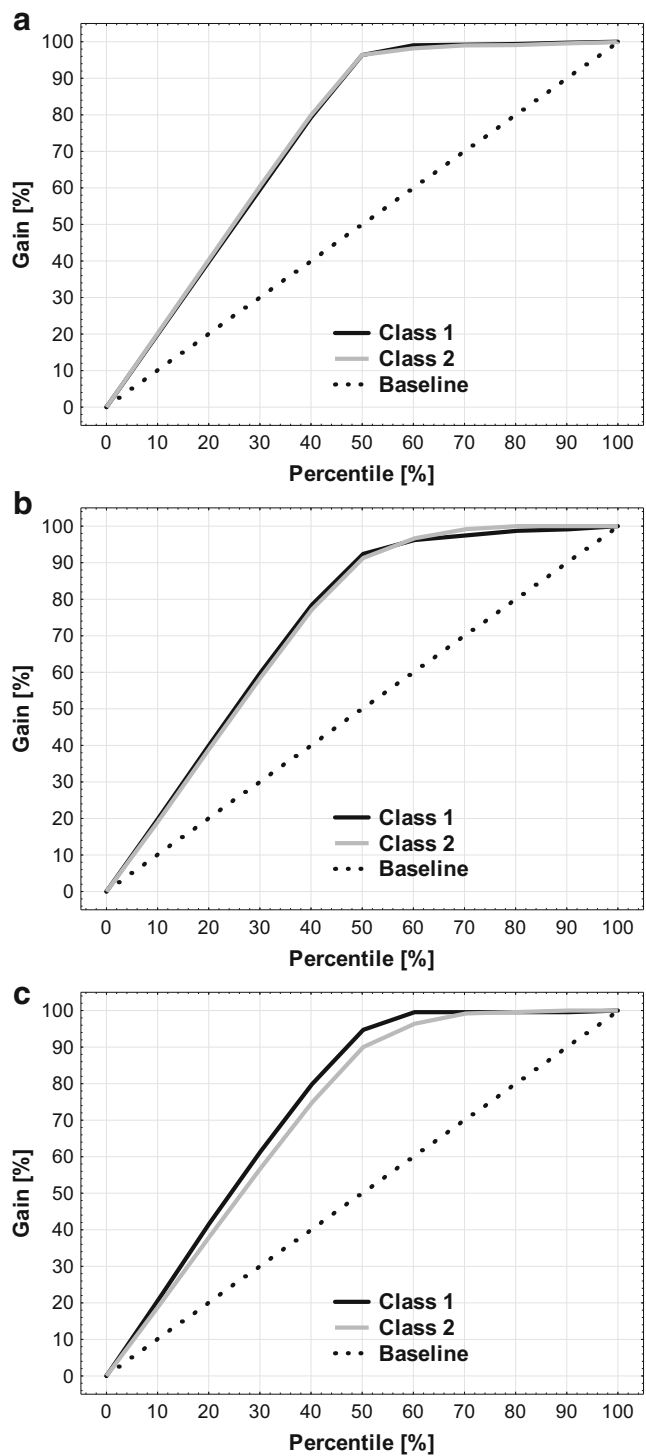

**Fig. 2** Cumulative gain charts for training (**a**), validation (**b**), and test sets (**c**) of MLP 100-19-2 network developed for criterion I classification system

86, PCB 83, PCB 114), six (PCB 139, PCB 147), seven (PCB 189, PCB 178), and nine (PCB 206) chlorine atoms. The majority of them were properly classified by all networks. According to criterion I, most of these compounds belong to class 1, which means that they do not easily undergo fragmentation. As it was mentioned, this behavior is typical for $\pi$-conjugated aromatic systems. Noteworthy, the high stability of PCBs and hence long half-life times is closely related to their persistence in the environment (Robertson and Hansen 2001; Hens and Hens 2017). Another groups of pollutants are pesticides and insecticides (oxychlordane, endrin, heptachlor). Interestingly, these compounds are characterized by very low or even zero molecular peak intensities (online source S4, Table S4), suggesting fast fragmentation (class 2). Another interesting examples of the class 2 are acid chlorides. The low stability of these compounds, which can be attributed to the presence of highly reactive (C=O)Cl group, does not exclude their significant impact on the environment. Noteworthy, toxic activity of these compounds on the aqueous organisms was well documented (Nabholz et al. 1993). Several acid chlorides can be found in the test set including 2-propenoyl chloride, 3-methyl-butanoyl chloride, octanoyl chloride, and 2-ethylhexanoyl chloride. All of them were properly classified by all models. Interestingly, according to the second criterion, these compounds belong to class 1 (Table S5), which means that the intensities of their [M-35] peaks are high. This suggest that the abstraction of chlorine atom proceeds rapidly. An interesting group of chloroorganics are also chlorinated aliphatic compounds. Several examples found in the test set are ethyl chloride, 5-chloropent-1-ene, 2,3-dichlorobutane, and 3-chloro-3-methyl-pentane. When analyzing criterion I-based models, chlorinated aliphatics are generally well classified by most of ANNs.

In order to evaluate the impact of each descriptor on the accuracy of the models, sensitivity analysis was performed. When considering molecular peak classification models (criterion I), three of the most important variables (online resource S3, Table S2) are atom type electrotopological state (E-state) descriptors, minaasC, nsssN, and maxdO developed by Hall and Kier (Hall and Kier 1995; Gramatica et al. 2000; Liu et al. 2001). These indices express minimum E-state value on aasC atom types, the number of sssN atoms in the molecule, and maximum E-state values on dO atoms, respectively. Another parameters of a high significance are C2SP2 (carbon type descriptor corresponding to $sp^2$ carbon atom attached to two other carbon atoms), path counts indices, piPC8 and piPC9 (Todeschini and Consonni 2009) and E-state parameters maxaasC, maxsssCH, maxaaCH, and minaaCH. Noteworthy, most of the parameters found among ten the most important, namely, minaasC, C2SP2, piPC8, piPC9, maxaasC, maxsssCH, maxaaCH, and minaaCH, are related to carbon atoms features and π-conjugation. The appearance of these molecular indices seems to be directly related to the stability of molecular peak. As it was mentioned, chlorinated aromatic hydrocarbons analogues such as PCBs, are less susceptible for fragmentation than aliphatic ones. This observation was confirmed by previous studies and can be explained by high stability of π-conjugated systems (Mohler et al. 1958; Sharma 2007; Nicolescu 2017). The role of particular descriptors in non-linear model is often not straightforward and easy to interpret. Nevertheless, some information can be inferred from their distributions. On Fig. 3, the box plots of ten of the most important variables, according to the sensitive analysis were presented. Interestingly, as evidenced by the parametric $T$ test and non-parametric Mann-Whitney $U$ and Kolmogorov–Smirnov tests ($p < 0.05$), the statistically important differences in distributions were observed for all descriptors except nsssN. This is of course a rough description. However, it shows that simple analysis of a particular variable regarded separately from the rest of parameters may be misleading, since according to the sensitive analysis, nsssN is ranked as the second most important variable (online resource S3, Table S2). Nevertheless, the good separation of classes 1 and 2 can be observed for other descriptors (Fig. 3). As it can be inferred, minaasC values are generally higher in case of compounds belonging to class 1. Since the highest minaasC values correspond to polychlorinated aromatic compounds, this seems to be consistent with the previously observed high intensity of PCBs' molecular peaks. The high stability of molecular ions containing several chlorine atoms can be explained by effective delocalization of unpaired electron on chlorine substituents attached to hydrocarbon π-conjugated systems. In general, the effect of resonance stabilization of molecular ion and characteristic for aromatic compounds can be illustrated by C2SP2 descriptor analysis. The highest C2SP2 was observed for compounds containing several aromatic rings. Some examples are tris(3-chlorophenyl)phosphine, chlorophacinone, and 2-chloro-1,4-dibenzamidobenzene. As it can be expected, compounds belonging to class 1 generally exhibit higher values of C2SP2 (Fig. 3). Another interesting descriptor is maxdO. In most cases, this parameter takes higher values for class 2 indicating fast fragmentation. Therefore, it can be considered as molecular ion instability measure. The maxdO descriptor is high for compounds containing relatively reactive carbonyl groups such as ketones, amides, and esters. On the other hand, it takes zero value for compounds containing no oxygen atoms. Noteworthy, molecular ions of esters and ketones are known to fragmentate readily via many paths such as inductive cleavage of the C–C bond next to carbonyl group, McLafferty rearrangement, or carbon monoxide elimination (Demarque et al. 2016).

Although classification models based on criterion II are less accurate, they can be useful for additional fragmentation behavior analysis. Noteworthy, many studies showed that the appearance of [M-35] peak on the spectra corresponding to the abstraction of chlorine atom from molecular ion is sensitive to the molecular structure features (Smith et al. 1972, 1973; Levy and Oswald 1976; Xu et al. 2000). The inspection of Table S3 (Supplementary material S3) shows that ten of the most important descriptors are atom type E-state indices (maxHaaCH, maxwHBd, maxHCHnX, nHCsatu, minHCsats, and nHBAcc) (Hall and Kier 1995; Gramatica et al. 2000; Liu et al. 2001), Barysz matrix descriptors (VE1_Dzm and VE1_DzZ) (Todeschini and Consonni 2009), one extended topochemical atom index (ETA_Shape_Y) (Roy and Ghosh 2004; Roy and Das 2011), and one topological charge descriptor (GGI8) (Todeschini and Consonni 2009). Similarly as in the case of criterion I-based model, descriptors related to carbon atom features and aliphatic/aromatic character can be also found in the criterion II-based model. Several of them, namely, maxHaaCH, maxHCHnX, nHCsatu, and minHCsats, were highly ranked by the sensitivity analysis. Other less important molecular indices are carbon types (C2SP2, C1SP2, C1SP3) and path counts indices (piPC8, piPC9, piPC10) (Todeschini and Consonni 2009).

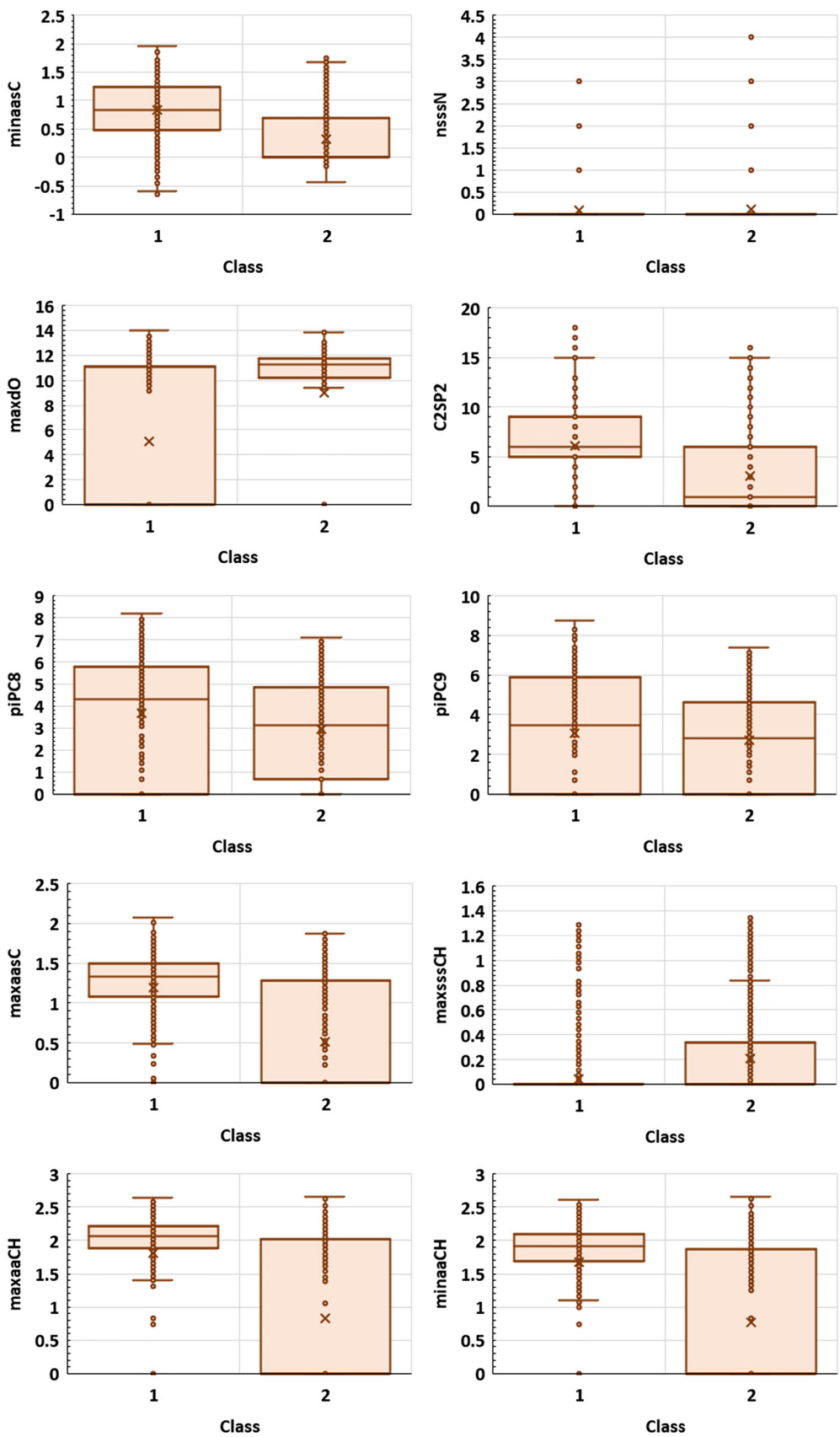

**Fig. 3** The distribution of the most important descriptors appeared in the criterion I-based model

**Table 2** Classification of selected MS spectra of sunscreens degradation and chlorination products performed using MLP 100-19-2 (model 1), MLP 100-23-2 (model 2), MLP 100-15-2 (model 3), MLP 100-25-2 (model 4) and MLP 100-21-2 (model 5)

| No. | Proposed compound | [M] | Class (exp.) | Source | Class (calc.) 1 | 2 | 3 | 4 | 5 |
|---|---|---|---|---|---|---|---|---|---|
| 1 | 2-Ethylhexyl 3,5-dichloro-4-(dimethylamino)benzoate, SMILES: CCCCC(CC)COC(=O)C1=CC(=C(N(C)C)C(=C1)Cl)Cl | 430 | 2 | (Sakkas et al. 2003) | 2 | 2 | 1 | 2 | 1 |
| 2 | 2-Ethylhexyl 3-chloro-4-(methylamino)benzoate, SMILES: CCCCC(CC)COC(=O)C1=CC=C(NC)C(Cl)=C1 | 571 | 2 | (Sakkas et al. 2003) | 2 | 2 | 2 | 2 | 2 |
| 3 | 2-Ethylhexyl 3,5-dichloro-4-(methylamino)benzoate, SMILES: CCCCC(CC)COC(=O)C1=CC(Cl)=C(NC)C(Cl)=C1 | 761 | 2 | (Sakkas et al. 2003) | 2 | 2 | 2 | 2 | 2 |
| 4 | 2-Ethylhexyl 4-amino-3-chlorobenzoate, SMILES: CCCCC(CC)COC(=O)C1=CC=C(N)C(Cl)=C1 | 430 | 2 | (Sakkas et al. 2003) | 2 | 2 | 2 | 2 | 2 |
| 5 | 2-Ethylhexyl 4-amino-3,5-dichlorobenzoate, SMILES: CCCCC(CC)COC(=O)C1=CC(Cl)=C(N)C(Cl)=C1 | 538 | 2 | (Sakkas et al. 2003) | 2 | 2 | 2 | 2 | 2 |
| 6 | 2-Ethylhexyl (2E)-3-(3-chloro-4-methoxyphenyl)prop-2-enoate, SMILES: CCCCC(CC)COC(=O)\C=C\C1=CC=C(OC)C(Cl)=C1 | 1099 | 1 | (Gackowska et al. 2016) | 2 | 2 | 2 | 2 | 2 |
| 7 | 2-Ethylhexyl (2E)-3-(3,5-dichloro-4-methoxyphenyl)prop-2-enoate, SMILES: CCCCC(CC)COC(=O)\C=C\C1=CC(Cl)=C(OC)C(Cl)=C1 | 68 | 2 | (Gackowska et al. 2016) | 2 | 2 | 2 | 2 | 2 |
| 8 | 3-chloro-4-methoxycinnamic acid, SMILES: COC1=C(C=C(C=C1)C(=O)O)Cl | 9999 | 1 | (Gackowska et al. 2014) | 1 | 1 | 1 | 1 | 1 |
| 9 | 3-chloro-4-methoxybenzaldehyde, SMILES: COC1=C(Cl)C=C(C=O)C=C1 | 9999 | 1 | (Gackowska et al. 2014) | 1 | 1 | 1 | 1 | 1 |
| 10 | 3,5-dichloro-4-methoxybenzaldehyde, SMILES: COC1=C(C=C(C=C1Cl)C=O)Cl | 9999 | 1 | (Gackowska et al. 2014) | 1 | 1 | 1 | 1 | 1 |
| 11 | 3-chloro-4-methoxyphenol, SMILES: COC1=C(C=C(C=C1)O)Cl | 7079 | 1 | (Gackowska et al. 2014) | 1 | 1 | 1 | 1 | 1 |
| 12 | 2,5-dichloro-4-methoxyphenol, SMILES: COC1=C(C=C(C(=C1)Cl)O)Cl | 5599 | 1 | (Gackowska et al. 2014) | 1 | 1 | 1 | 1 | 1 |
| 13 | 1-Chloro-4-methoxybenzene, SMILES: COC1=CC=C(C=C1)Cl | 9999 | 1 | (Gackowska et al. 2016) | 1 | 1 | 1 | 1 | 1 |
| 14 | 1,3-Dichloro-2-methoxybenzene, SMILES: COC1=C(C=CC=C1Cl)Cl | 9499 | 1 | (Gackowska et al. 2016) | 1 | 1 | 1 | 1 | 1 |
| 15 | 2-Ethylhexyl chloroacetate, SMILES: CCCCC(CC)COC(=O)CCl | 0 | 2 | (Gackowska et al. 2016) | 2 | 2 | 2 | 2 | 2 |
| 16 | 2,4-Dichlorophenole, SMILES: C1=CC(=C(C=C1Cl)Cl)O | 9999 | 1 | (Gackowska et al. 2016) | 1 | 1 | 1 | 1 | 1 |
| 17 | 2,6-Dichlor-1,4-benzoquinone, SMILES: C1=C(C(=O)C(=CC1=O)Cl)Cl | 7699 | 1 | (Gackowska et al. 2016) | 1 | 1 | 1 | 1 | 1 |
| 18 | 1,2,4-Trichloro-3-methoxybenzene, SMILES: COC1=C(C=CC(=C1Cl)Cl)Cl | 6199 | 1 | (Gackowska et al. 2016) | 1 | 1 | 1 | 1 | 1 |
| 19 | 2,4,6-Trichlorophenole, SMILES: C1=C(C=C(C(=C1Cl)O)Cl)Cl | 9999 | 1 | (Gackowska et al. 2016) | 1 | 1 | 1 | 1 | 1 |
| 20 | 3,5-Dichloro-2-hydroxyacetophenone, SMILES: OC1=C(Cl)C=C(Cl)C=C1 | 769 | 2 | (Gackowska et al. 2016) | 1 | 1 | 1 | 1 | 1 |
| 21 | 2-chloro-1-(4-methoxyphenyl)ethan-1-one, SMILES: COC1=CC=C(C=C1)C(=O)CCl | 851 | 1 | (Kalister et al. 2016) | 1 | 1 | 1 | 1 | 1 |
| 22 | 1-(4-t-butylphenyl)-2-chloro-3-(4-methoxyphenyl)propane-1,3-dione, SMILES: COC1=CC=C(C=C1)C(=O)C(Cl)C(=O)C1=CC=C(C=C1)C(C)(C)C | 194 | 2 | (Trebše et al. 2016) | 2 | 2 | 2 | 2 | 2 |
| 23 | 1-(4-t-butylphenyl)-2,2-dichloro-3-(4-methoxyphenyl)propane-1,3-dione, SMILES: COC1=CC=C(C=C1)C(=O)C(Cl)(Cl)C(=O)C1=CC=C(C=C1)C(C)(C)C | 0 | 2 | (Trebše et al. 2016) | 2 | 2 | 2 | 2 | 2 |
| 24 | 2-benzoyl-4-chloro-5-methoxyphenol, SMILES: COC1=CC(O)=C(C=C1Cl)C(=O)C1=CC=CC=C1 | 1515 | 1 | (Zhang et al. 2016) | 1 | 1 | 1 | 1 | 1 |
| 25 | 6-benzoyl-2,4-dichloro-3-methoxyphenol, SMILES: COC1=C(Cl)C(O)=C(C=C1Cl)C(=O)C1=CC=CC=C1 | 1512 | 1 | (Zhang et al. 2016) | 1 | 1 | 1 | 1 | 1 |
| 26 | 2,4,6-trichloro-3-methoxyphenol, SMILES: COC1=C(Cl)C(O)=C(Cl)C=C1Cl | 1000 | 1 | (Zhang et al. 2016) | 1 | 1 | 1 | 1 | 1 |

### Exemplary application of models

In our previous works (Gackowska et al. 2014, 2016; Studziński et al. 2017), degradation of popular UV filters in the presence of different oxidizing and chlorinating agents was studied. Sunscreen agent contamination deserves special attention, due to the widespread use of organic UV filters in personal care products (Santos et al. 2012). Furthermore, these compound are relatively stable and therefore resistant to the wastewater treatment (Ramos et al. 2015, 2016). In this section, mass spectra of several sunscreen agents, 2-ethylhexyl-4-methoxycinnamate (EHMC), 2-ethylhexyl 4-(dimethylamino)benzoate (ODPABA), avobenzone, and oxybenzone chlorination by-products were analyzed. Due to the large variety of detected compounds, these results can be useful for additional validation of proposed classification networks. Presented in Table 2, data comprises molecular peaks intensities reported by our group and by other authors. In order to apply the proposed classification criterion, the MS peak intensities were scaled to a NIST units. In some cases, the intensity values were obtained from graphic data. This can be easily done using ImageJ (Schneider et al. 2012), which is a comprehensive software dedicated for image analysis.

As one can see from Table 2, the majority of EI-MS spectra belonging to the class 1 correspond to aromatic compounds with chlorinated phenyl ring. However, the presence of aromatic moiety does not always indicate the appearance of high molecular peak on the MS spectra. In several cases, including aromatic compounds (2-ethylhexyl 3,5-dichloro-4-(dimethylamino)benzoate, 2-ethylhexyl 4-amino-3-chlorobenzoate, 2-ethylhexyl (2E)-3-(3,5-dichloro-4-methoxyphenyl)prop-2-enoate, 2-ethylhexyl chloroacetate, 1-(4-*t*-butylphenyl)-2-chloro-3-(4-methoxyphenyl)propane-1,3-dione, 1-(4-*t*-butylphenyl)-2,2-dichloro-3-(4-methoxyphenyl)propane-1,3-dione), the intensity of molecular peak is very low (Table 2). This can be caused by the steric hindrance effect which have been already described. The lack of molecular peaks may cause some difficulties in degradation product identification. Fortunately, most of these compounds were properly classified. Interestingly, in case of 2-ethylhexyl 3,5-dichloro-4-(dimethylamino)benzoate, two proposed models, MLP 100-15-2 and MLP 100-21-2, failed. This shows that all five networks should be taken into account when analyzing EI-MS spectra. As one can see form Table 2, there are only two spectra wrongly classified by all models, namely, 2-ethylhexyl (2E)-3-(3-chloro-4-methoxyphenyl)prop-2-enoate and 3,5-dichloro-2-hydroxyacetophenone. However, in case of 3,5-dichloro-2-hydroxyacetophenone which was assigned to the class 1, the intensity of molecular peak was slightly lower than classification threshold (800 NIST units). In such cases, it is difficult to unambiguously assign compounds, since depending on the EI-MS spectra recording conditions, slightly different peak intensities may be obtained. Another example of molecular peak close to 800 NIST units can be observed for 2-chloro-1-(4-methoxyphenyl)ethan-1-one. Fortunately, this compound was properly assigned to class 1. It is worth to note that, there is only one false-positive example of class 1 (2-ethylhexyl (2E)-3-(3-chloro-4-methoxyphenyl)prop-2-enoate). The intensity of molecular peak of this 2-ethylhexyl-4-methoxycinnamate (EHMC) chlorinated disinfection by-product is 2500, which means that it should not be classified to class 2.

### Conclusions

Since simple EI-MS approach is still one of the most commonly used methods in pollutant environmental monitoring, it is important to develop theoretical tools of MS spectra interpretation. Detection of new compounds is often problematic due to the lack of analytical standards and reference spectra in the MS databases. However, there are many rules of molecular ion fragmentation, which can be helpful in MS spectra analysis. These rules are based on the structural features of the molecules. For instance, there are characteristic fragmentation pathways of aldehydes, esters, amines, etc. The rapid development of QSPR methods allowing for the support of chemical compounds identification was mainly focused on the retention parameters modelling (Katritzky et al. 2000; Kaliszan 2007). However, several attempts of MS spectra modelling appeared in the literature. Two major approaches can be distinguished, namely, predicting MS spectra features using quantum-chemical computations (Cautereels et al. 2016; Ásgeirsson et al. 2017; Spackman et al. 2018) and 2D structure and topology-based methods (Gray et al. 1980; Gasteiger et al. 1992; Copeland et al. 2012). The latter approach can be regarded as an extension of popular fragmentation rules. The similar concept was presented in this paper. We have investigated the applicability of chlorinated compounds MS spectra classification model based on the 1D and 2D molecular descriptors. The mass spectra were classified based on the two characteristic [M] and [M-35] peak intensities. However the first criterion due to the high accuracy of prediction was found to be more appropriate for analytical purposes. Apart from the standard validation procedure, the selected models were tested against some additional examples of chlorinated compounds spectra reported in the literature. The majority of these spectra were properly classified by all networks. This shows that the approach presented in this study can be helpful for the identification of unknown chlorinated compounds. Although the models does not generate the structure form the spectra, they can be useful for confirmation of the hypothetical structure by checking whether the theoretical classification of the potential candidate meets the experimental results. It is worth to emphasize that in this study, only simple descriptors based on the 1D and 2D structure were taken into account. Therefore, the presented approach can be probably developed by using more advanced descriptors or dividing population into more than two classes. Therefore, it seems to be reasonable to focus on the further development of mass spectral prediction methods based on neural networks and molecular descriptors.